\documentclass[12pt]{article}  
\usepackage[centertags]{amsmath}
\usepackage{color}
\usepackage{amsfonts}
\usepackage{amssymb}
\usepackage{amsthm}
\usepackage{newlfont}

\def\dalpha{{\dot\alpha}}
\def\dbeta{{\dot\beta}}

\def\oeta{{\overline{\eta}}}

\begin{document}

\title{Two-Twistor Space, \\ Commuting Composite Minkowski Coordinates
\\ and Particle Dynamics}

\author{A. Bette$^{1)}$ , J. Lukierski$^{2)}$ and C. Miquel-Espanya$^{3)}$ 
\\ 
  \\
 $^{1)}$ Royal Institute of Technology \\
 S-151 81 S\"{o}dert\"{a}lje, Sweden
 \\  
 $^{2)}$Institute for Theoretical Physics, \\ University of Wroc\l{}aw, \\
 pl. M. Borna 9, 50-204 Wroc{\l}aw, Poland
  \\  
$^{3)}$Departament de F\'{\i}sica Te\`{o}rica i IFIC, \\ Universitat de Val\`{e}ncia, \\
 Dr. Moliner 50, 46100 Burjassot (Val\`{e}ncia), Spain
  }
\date{}
\maketitle   

\begin{abstract}
We employ the  modification of the basic Penrose formula in twistor
theory, which allows to introduce commuting composite space-time
coordinates. It appears that in the course of such modification the
internal symmetry $SU(2)$ of two-twistor system is broken to $U(1)$.
We consider the symplectic form on two-twistor space, permitting to
interpret its 16 real components as a phase-space. After a suitable
change of variables such a two-twistor phase space is split into
three mutually commuting parts, describing respectively the standard
relativistic phase space (8 degrees of freedom), the spin sector (6
degrees of freedom) and the canonical pair angle-charge describing
the electric charge sector (2 degrees of freedom). We obtain a
geometric framework providing a twistor-inspired 18-dimensional
extended relativistic phase space $\mathcal{M}^{18}$. In such a
space we propose the action only with first class constraints,
describing the relativistic particle characterized by mass, spin and
electric charge.
\end{abstract}


\section{Introduction}

$\quad\,$The choice of basic geometric variables that describe the
dynamics at the most elementary level is an important issue
extensively discussed in mathematical physics as well as in
fundamental interactions theory. In standard relativistic $D=4$
theory we assume that the basic geometry is described by the
Minkowski space-time coordinates $x_\mu = (\overrightarrow{x},
x_{0}=c\, t)$.

There are two ways of extending the notion of  classical Minkowski
space-time:

i) One adds additional geometric degrees of freedom,
 e.g. the anticommuting Grassman variables
in supersymmetric theory or additional commuting continuous or
discrete coordinates. In principle the replacement of elementary
point particles by strings can be also described as the extension of
Minkowski space by infinite set of auxiliary coordinates describing
Fourier modes of an extended object. In all these approaches the
space-time coordinates remain elementary.

ii) One can consider the space-time geometry as a derived notion,
with composite space-time coordinates. Because the most elementary
representation of the Lorentz algebra is spinorial\footnote{By most
elementary representation we mean that all other irreducible
representation can be obtained by tensoring procedure.}, natural
candidates for new elementary coordinates are Lorentz spinors.
Taking into consideration that the mass can be considered as a
dynamical effect, these elementary spinorial coordinates describing
primary kinematics should describe the geometry of
 massless world with basic conformal
invariance. In such a way we arrive at the notion of twistors
 (see e.g. [1--5]) - fundamental representations
of the conformal algebra - as describing the coordinates of primary
geometry.

In four dimensions ($D=4$) the conformal algebra is
 $SU(2,2)=\overline{SO(4,2)}$, and its
fundamental twistor representation $T^4 = (Z_1, Z_2, Z_3, Z_4)$ is
complex. We define the twistor space as a fourdimensional complex
metric space $T^4 = (c^4, h)$, where the Hermitean metric $h$ has
the signature
 $(+,+,-,-)$.
 Choosing
 \begin{equation}\label{betlu1}
    h = \begin{pmatrix} 0 & I_2 \cr
    I_2 & 0
    \end{pmatrix} \, , 
\end{equation}
one can represent a twistor by a pair of $D=4$ Weyl
spinors\footnote{We define $\omega^\alpha = \epsilon^{\alpha
\beta}\omega_{\beta}$, $\overline{\omega}^{\dot{\alpha}} =
\epsilon^{\dot{\alpha}\dot{\beta}}\overline{\omega}_{\dot{\beta}}$,
$\omega_\alpha=\omega^\beta\epsilon_{\beta\alpha}$,
$\overline{\omega}_{\dot\alpha}=\overline{\omega}^{\dot\beta}\epsilon_{\dot\beta\dot\alpha}$
and $\epsilon^{\alpha\beta}\epsilon_{\beta\gamma} =- \delta_{\
\gamma}^{\alpha}$,
$\epsilon^{\dot{\alpha}\dot{\beta}}\epsilon_{\dot{\beta}\dot{\gamma}}
 = -\delta_{\dot\gamma}^{\dot\alpha}$, where $\epsilon^{\alpha\beta}
 =\epsilon_{{\alpha}{\beta}} =
  \begin{pmatrix} 0 & - 1 \cr 1 & 0
    \end{pmatrix}
    = - \epsilon^{\dot\alpha\dot\beta} = - \epsilon_{\dot{\alpha}\dot{\beta}}$. }
 ($\alpha, \beta = 1,2 $; $A= 1,2,3,4$)
\begin{equation}\label{betlu2}
    Z_A = \left(
    \omega^\alpha, \overline{\pi}_{\dot{\beta}}
    \right)\, ,
    \qquad
    \overline{Z}_{A} = (Z_A)^{*} = \left(\overline{\omega}^{\dot{\alpha}} ,
    \pi_{\beta} \right) \, .
\end{equation}
More explicitly, the $SU(2,2)$ norm can be written as follows
\begin{equation}\label{betlu3}
    \langle T, T\rangle =
 Z_A \, h^{A B} \, \overline{Z}_{B}
= Z_A \, \overline{Z}^{A} =
\omega^\alpha\pi_\alpha+\overline{\omega}^{\dot\alpha}\overline{\pi}_{\dot\alpha}\,
.
\end{equation}

The link with space-time coordinates is obtained by imposing the
Penrose incidence relations
\begin{equation}\label{betlu4}
 \omega^{{\alpha}} =   i \, z ^{\alpha\dot\beta} \,
    \overline{\pi}_{\dot\beta}\,,
\quad
    \overline{\omega}^{\dot{\alpha}} =  -i \, \overline{z} ^{{\beta}\dot{\alpha}} \,
    {\pi}_{{\beta}}\,,
\end{equation}
where
\begin{equation}\label{betlu5}
z ^{{\alpha}\dot{\beta}} =
{\textstyle\frac{1}{2}}(\sigma_\mu)^{\alpha \dot{\beta}}\, z^\mu \,
, \qquad \overline{z}^{\beta\dot{\alpha}} = {\textstyle\frac{1}{2}}
(\overline{\sigma}_{\mu})^{\beta\dot{\alpha}}\ \overline{z}^\mu \, ,
\end{equation}
describe complex Minkowski coordinates. {}From (\ref{betlu4})
follows that the complex Minkowski coordinates parametrize
two-planes in $T^4$, i.e.

\begin{equation}\label{betlu6}
    z_\mu \in G_{4;2} (c) =
    \frac{SU(2,2)}{S(U(2) \times U(2))} \, .
\end{equation}
Any such plane is parametrized by a pair of nonparallel twistors
\begin{equation}\label{betlu7}
    Z_{A; i} = \left( \omega^{\alpha}_{\ \ ; i}, \overline{\pi}_{\dot{\beta}; i}
    \right)
    \quad
      \overline{Z}_{A}^{ \ \ ;i}
= \left( Z_{A; i} \right)^{*} =
       \left( \overline{\omega}^{\dot{\alpha}; i},
       {\pi}_{{\beta}}^{\ ;i}
    \right)  \qquad i=1,2 \
    \, ,
\end{equation}
where we observe that the complex-conjugated spinors are
contravariant in the internal $U(2)$ index space, i.e. $(\pi_{\alpha
; i})^{*} = \overline{\pi}_{\dot{\alpha}}^{\ \ ; i}$ etc. In such a
way the
 $U(2)$-invariant norm we denote
 $A_{; i}(A_{; i})^{*} = A_{; i}\overline{A}^{; i}$.\footnote{
We define by analogy with the footnote 2 that
$A^{;i}=\epsilon^{ij}A_{;j}$, $A_{;i} = A^{;j}\epsilon_{ji}$ where
$\epsilon^{ij}=\begin{pmatrix} 0 &-1 \cr 1 &0
\end{pmatrix}=\epsilon_{ij}$. 
}
Writing down the relation (\ref{betlu4}) for two twistors

\begin{equation}\label{betlu8}
\omega^{{\alpha}}_{\ \ ; i}  = i \ z^{{\alpha}\dot\beta}\
\overline{\pi}_{\dot\beta;i}\, , \qquad
\overline{\omega}^{\dot{\alpha} ; i} = - i \
\overline{z}^{{\beta}\dot{\alpha}}\ {\pi}_{\beta}^{\ ;i}\, ,
\end{equation}
one gets as the solution of (\ref{betlu8}) the \underline{complex}
 composite Minkowski space-time coordinates
 \begin{equation}\label{betlu9}
    z^{{\alpha}\dot\beta} =
    x^{{\alpha}\dot{\beta}} + i \, y^{{\alpha}\dot{\beta}}
    =
    \frac{i}{f} \, \omega^{{\alpha}; i}
     \, \overline{\pi}^{\dot\beta}_{\ \ ; i}\, ,
 \end{equation}
where $A^{ ; i } = \epsilon^{ij}\, A_{ ; j}$, the fourvectors
 $x_\mu , y_\mu$  are real, and
 \begin{equation}\label{9a}
    f =
\overline{\pi}^{\dot\alpha}_{\phantom{\alpha}; 1} \,
\overline{\pi}_{\dot\alpha ; 2 }= \textstyle\frac{1}{2}
    \overline{\pi}_{\dot\alpha ; i} \, \overline{\pi}^{\dot\alpha}_{\phantom{\alpha}; j }\
          \epsilon^{ij} =
    \textstyle\frac{1}{2}\, \overline{\pi}_{\dot\alpha ; i}\, \overline{\pi}^{\dot\alpha ; i}
\, , \qquad \overline{f} = \textstyle\frac{1}{2} {\pi}_{{\alpha};i}
 \, \pi^{{\alpha}; i }\ ,
\end{equation}
consistently with the numerical equality $\epsilon_{ij} =
\epsilon^{ij}$. From the relation (\ref{betlu9}) one can show that
in order to embed the real Minkowski  coordinates in twistor spaces
one should consider pairs of twistors $T_1, T_2$ which span a null
2-plane, i.e.
\begin{equation}\label{betlu10}
    t_{i}^{\ j} = \langle T_i, T_j \rangle
 = Z_{A;i} \, h^{AB} \, \overline{Z}_{B;} ^{\phantom{B} j}
     = 0 \, .
\end{equation}
In such a case one gets from (\ref{betlu8}) that $y_\mu = 0$, i.e.
the formula (\ref{betlu9}) describes \underline{real} Minkowski
coordinates.

The Hermitean metric $h$ generates the $SU(2,2)$-invariant
symplectic two-form $\Omega$. Using (\ref{betlu1}) and
(\ref{betlu7}) one can write
\begin{eqnarray}\label{betlu11}
  \Omega &=&
 i \, d \, Z_{A; i} \wedge  d \overline{Z}_{B}^{\phantom{B}; i} \,
h^{AB} \cr
   &=&
i\left( d\, {\omega}^{\alpha}_{\phantom{\alpha}; i} \, \wedge \, d
\, {\pi}_{\alpha}^{\phantom{\alpha}; i} + d \,
\overline{\pi}_{\dot{\alpha} ; i}  \, \wedge d \,
\overline{\omega}^{\dot{\alpha}; i} \right)
\end{eqnarray}
and the Liouville one form $\Theta$, satisfying the relation $\Omega
= d \Theta$, looks as follows

\begin{equation}\label{betlu12}
    \Theta =   \,
{\textstyle\frac{i}{2}} \left(
 {\omega}^{\alpha}_{\phantom{\alpha}; i}  \, d \pi_{\alpha}^{\phantom{\alpha}; i}
+ \overline{\pi}_{\dot{\alpha}; i}
 \, d \overline{\omega}^{\dot{\alpha}; i}
- H.C \right)\, .
 \end{equation}
It appears that the two-form (\ref{betlu11}) defines fundamental
twistorial Poisson brackets (TPB) and the Liouville one-form
(\ref{betlu12})
 should be important for the construction of dynamical Lagrangean and Hamiltonian
models in twistor space.

The problem which we firstly consider in Sect. 2 is the proper
notion of composite space-time in twistor space. If we use the
formula (\ref{betlu9}) it was shown \cite{betl6} that the real
composite Minkowski space coordinates
 $x_\mu = Re \, z_\mu$ are noncommuting.
Following \cite{betl9,betl10,betl11} we introduce the modification
of the standard Penrose formula (\ref{betlu9}) which leads to
commuting composite space-time coordinates $X_\mu$. Further in Sect.
\ref{extendedspace} by considering the nonlinear
 transformations of sixteen real coordinates in two-twistor space we introduce
an enlarged 18-dimensional relativistic phase space
$\mathcal{M}^{18}$ with three mutually commuting sectors

i) Relativistic phase-space $(X_\mu, P_\mu)$

ii) Spinorial complex phase space $(\eta_\alpha, \sigma_\alpha,
\overline{\eta}_{\dot{\alpha}}, \overline{\sigma}_{\dot{\alpha}})$

iii) Electric charge phase space $(e,\phi)$

One can prove that the equivalence with the two-twistor space
(\ref{betlu7}) implies the imposition of two constraints in the
spinorial complex phase space \cite{betl10new}.

In Sect. \ref{relativisticparticles} we consider in the extended
phase space $\mathcal{M}^{18}=(X_\mu,P_\mu,\eta_\alpha,
\overline{\eta}_{\dot\alpha}, \sigma^{\alpha},\\
\overline{\sigma}_{\dot\alpha} ,e,\phi)$ a free particle model. In
\cite{betl10new} we considered such a model with two geometric
second class constraints, and three physical
 first class constraints defining physical quantities:
mass, spin and charge. It appears however that the classical Dirac brackets, obtained in the
process of elimination of the second class constraints, are very difficult to quantize in a
consistent way (Jacobi identities!). In this paper we propose an alternative model in
$\mathcal{M}^{18}$, with the two second class constraints proposed in \cite{betl10new} replaced by
one geometric first class constraints. In such a model with four first class constraints (one
geometric, three physical) the quantization is straightforward.

\message{In Sect. 4 we describe in our extended phase space the
Hamiltonian formalism and by considering the electromagnetic
coupling we justify that the two-dimensional phase space $(e,\phi)$
can be interpreted  as electric charge and its dually conjugated
 angle  coordinate.}

In the last Section we shall present the concluding remarks.

\section{Standard relativistic phase space from two-twistor
 geometry}

$\quad\,$ The two-twistor symplectic form (\ref{betlu11}) implies
the
 following nonzero fundamental Poisson brackets
 \begin{eqnarray}\label{betlu13}
   \{ \pi_{\alpha}^{\phantom{\alpha}; i}, {\omega}^{{\beta}}_{\phantom{\beta}; j} \} &=&
 i \, \delta_{\alpha}^{\beta} \, \delta^{i}_j \, ,
    \cr\cr
     \{ \overline{\pi}_{\dot{\alpha}; i},
      {\overline{\omega}}^{\dot{\beta};j} \} &=&
 -  i \, \delta_{\dot{\alpha}}^{\dot{\beta}} \, \delta_{i}^{j} \, .
 \end{eqnarray}
 After quantization one obtains from (\ref{betlu13}) the basic
 twistorial canonical commutation relations (TCCR)
 \begin{eqnarray}\label{betlu14}
 [ \widehat{\omega}^{{{\beta}}}_{\phantom{\alpha}; j},
 \widehat{{\pi}}_{\alpha}^{\phantom{\alpha}; i} ]
 & = &
 \hbar \, \delta^{\beta}_\alpha \, \delta^i_j \, ,
 \cr\cr
 [ \widehat{\overline{\omega}}^{{\dot\beta};j},
 \widehat{{\overline{\pi}}}_{\dot{\alpha}; i} ]
 & = &
 - \hbar \, \delta_{\dot\alpha}^{\dot\beta} \, \delta_i^j\, .
 \end{eqnarray}
 Using the relation (\ref{betlu9}) one can calculate the TPB of
 the real composite Minkowski coordinates $x_\mu$. One gets
 \cite{betl6}

 \begin{equation}\label{betlu15}
    \{ x_\mu , x_\nu \} = - \frac{1}{m^4} \, \epsilon_{\mu\nu\rho\tau}
    \, W^{\rho}\, P^{\tau} \, ,
\end{equation}
where

\begin{equation}\label{16}
    m^2 =
 2\left|  f \right|^2 =
    2\left|\textstyle\frac{1}{2}  \overline{\pi}_{\dot\alpha; i} \, \overline{\pi}^{\dot\alpha; i} \right| ^2 =
    \textstyle\frac{1}{2}\left|\overline{\pi}_{\dot\alpha; i} \, \overline{\pi}^{\dot\alpha; i} \right|^2 =P_\mu \, P^\mu
    \, ,
\end{equation}
and $(r=1,2,3)$

\begin{eqnarray}\label{betlu17}
  P^{\alpha\dot{\beta}} &=&
  \pi^{\alpha ;i} \,
  \overline{\pi}^{\dot{\beta}}_{\phantom{\dot\beta}; i} \, ,
\qquad \quad \qquad  P^{\mu} = \sigma^\mu _{\phantom{\mu}\alpha
\dot{\beta}} \,
  P^{\alpha\dot{\beta}} \, ,
   \\
   \label{betlu18}
 W^{\alpha\dot{\beta}}
  &=&
 \pi^{\alpha ;i} (\tau^r )_{i}^{\phantom{i} j} \,
 \overline{\pi}^{\dot{\beta}}_{\phantom{\beta}; j} \, t_r
 \, ,
 \qquad
 W^\mu = \sigma^\mu _{\phantom{\mu} \alpha\dot{\beta}} \, W^{\alpha\dot{\beta}}\, ,
\end{eqnarray}
where $(a=0,1,2,3)$
\begin{equation}\label{betlu19}
    t_{i}^{\phantom{i} j} =
     (\tau^{a})_{i}^{\phantom{i} j}\, t_a = \langle T_i , T_j\rangle
    = Z_{A ; i} \, \overline{Z}^{A ; j}\, ,
\end{equation}
and $(\tau^0)_{i}^{\phantom{i} j} = \begin{pmatrix}1 & 0 \cr 0 &1
\end{pmatrix}$ and $(\tau^r)_{i}^{\phantom{i} j} $ describe three
Pauli isospin matrices. The internal isospin symmetry is represented
by the following $su(2) \otimes u(1)$ Poisson algebra brackets
$(r,s,u=1,2,3)$
\begin{equation}\label{betlu20}
    \{ t_r , t_s \} = \epsilon_{rsu}\, t_u \, ,
    \qquad
    \{ t_0 , t_r \} = 0 \, ,
\end{equation}
as it follows from  (\ref{betlu13}) and (\ref{betlu19}).

Using the relation (\ref{betlu17}) one can extended the TPB
(\ref{betlu15}) by the following two relations:

\begin{equation}\label{betlu21}
\{P_\mu, x_\nu \}  =  \eta_{\mu\nu} \, ,
\end{equation}
\begin{equation}\label{betlu22}
\{ P_\mu , P_\nu \}  =
  0 \, .
\end{equation}

Replacing the TPB (\ref{betlu13}) and (\ref{betlu21}--\ref{betlu22})
by TCCR (\ref{betlu14}) one gets the quantized relativistic phase
space $(\widehat{x}_\mu , \widehat{P}_{\mu})$ with noncommuting
composite Minkowski coordinates $\widehat{x}_\mu$. Such
noncommutativity in the presence of nonvanishing spin $(W_\mu \ne
0)$ can be traced back to earlier considerations by Souriau
\cite{betl7} and Casalbuoni \cite{betl8}. Indeed, the composite
four-vector (\ref{betlu18}) can be identified with the
Pauli-Lubanski vector in arbitrary relativistic frame. It is
orthogonal, as it should be, to the composite fourmomentum
 (\ref{betlu17})
 \begin{equation}\label{betlu23}
    P_\mu \, W^\mu = 0 \, ,
\end{equation}
and in the rest system\footnote{In quantized systems such
description implies the consideration of eigenstates of the
four-momentum operator $\widehat{P}_{\mu}$.} $P_\mu = (m,0,0,0)$ one
can write the noncommutativity relations of quantized composite
 Minkowski coordinates as folows:
  $(\widehat{x}_\mu = (\widehat{x}_k, \widehat{x}_{0} = c\, \widehat{t } )$
  \begin{equation}\label{betlu24}
    \left[ \widehat{x}_k , \widehat{x}_l \right] =
    - i \, \frac{\hbar}{m^2} \, \epsilon_{klm}\, \widehat{S}_{m}\, ,
\end{equation}
where $\widehat{W}_k = m\widehat{S}_k$ and as follows from
(\ref{betlu18})
\begin{equation}\label{betlu25}
    \left[ \widehat{S}_k , \widehat{S}_{l} \right]
     = i\, \hbar \, \epsilon_{klm}\, \widehat{S}_{m}\, .
\end{equation}

In this lecture we would like to consider the composite relativistic
phase space $(X_\mu, P_\mu)$, satisfying the standard TPB

\begin{eqnarray}\label{betlu26}
\{ X_\mu, X_\nu \} & = & \{ P_\mu , P_\nu \} = 0 \, , \cr\cr \{
X_\mu, P_\nu \} & = & \eta_{\mu \nu} \, .
\end{eqnarray}
In such a case one has to change the definition of composite
Minkowski coordinates by the modification of the standard definition
(\ref{betlu9}), which
 we redefine as follows \cite{betl9,betl10,betl11,betl10new}


\begin{eqnarray}\label{betlu27}
z^{\alpha\dot{\beta}} \longrightarrow Z^{\alpha\dot{\beta}} & = &
 z^{\alpha\dot{\beta}} + \Delta z^{\alpha\dot{\beta}}
 = X^{\alpha\dot{\beta}} + i \, Y ^{\alpha\dot{\beta}}
 \cr\cr
 & = &
 z^{\alpha\dot{\beta}} + i \,
 (t_1 - i \,  t_2) \ \frac{ \pi^{\alpha ;1} \, \overline{\pi}^{\dot{\beta}}_{\ \ ;  2}}
 { |f|^2} \, ,
\end{eqnarray}
or

\begin{eqnarray}\label{betlu28}
X^{\alpha\dot{\beta}} & = & x^{\alpha\dot{\beta}} +\Delta
x^{\alpha\dot{\beta}} = x^{\alpha\dot{\beta}} -
 \frac{1}{2\, |f|^2  }\,
 \left[
 t_1 \, \pi^{\alpha ; i} (\tau_2)_{i}^{\ j} \,
 \overline{\pi}^{\dot{\beta}}_{\ ;  j}
 - t_2 \, \pi^{\alpha ; i}
 (\tau_1)_{i}^{\ j} \, \overline{\pi}^{\dot{\beta}}_{\ ; j}
 \right]
 \cr\cr
 & = &
 x^{\alpha\dot{\beta}} -
  \frac{1}{2 | f |^2} \ \epsilon_{3rs}
  \, t_r \, \pi^{\alpha ; i}
  (\tau_s)_{i}^{\ j} \, \overline{\pi}^{\dot{\beta}}_{\ ; j}\  .
\end{eqnarray}
One can show that the TPB of the composite coordinates
(\ref{betlu28}) are given by the relation  (\ref{betlu26})
 i.e. after quantization we obtain commuting composite Minkowski
  coordinate. We see from the relations
 (\ref{betlu28}) that the commutative coordinates $X^\mu = (\sigma^\mu)
 _{\alpha \dot{\beta}} \, X^{\alpha \dot{\beta}}$ distinguish a third
 direction in the isospace $O(3) \simeq SU(2)$ i.e. break the
 isospin symmetry from $O(3)$ to $O(2)$.

\section{Extended  relativistic phase space from two-twistor
space.}\label{extendedspace}

$\quad\,$Let us consider the symplectic 2-form (\ref{betlu12}) and
insert the formula (\ref{betlu9}) for the composite complex
Minkowski
 coordinates. Using the $SU(2)$-covariant notation\footnote{The advantage of such
 a notation has been pointed out to us by S. Fedoruk.
 In such a framework  the ``half-twistors" $\pi_{\alpha ; i}$,
$\overline{\pi}_{\dot{\alpha}; i}$ can be treated as a pair of
$Sl(2;C)$ harmonic
 spinors (see e.g. \cite{betl12,betl13,betl14}) with particular normalization.}
  one gets
 \begin{equation}\label{betlu29}
 \Theta = \pi _{\alpha}^{\phantom{\alpha}; i }\,
 \overline{\pi}_{\dot{\beta}; i}
  \, d x^{\alpha \dot{\beta}}
 + i \,
 y^{\alpha \dot{\beta}}
 \left(
 \pi_{\alpha}^{\phantom{\alpha}; i} \, d \, \overline{\pi}_{\dot{\beta} ;i}
 - \overline{\pi}_{\dot{\beta} ; i} \, d \pi_{\alpha}^{\phantom{\alpha}; i}
 \right) \, .
\end{equation}
Using the formula for the imaginary part of the complex four-vector
$z_\mu$ one gets

\renewcommand{\theequation}{31\alph{equation}}
\setcounter{equation}{0}
\begin{equation}\label{betlu30a}
    t^{i}_{\phantom{i} j} = - 2 \, y^{\alpha \dot{\beta}} \,
    \pi_{\alpha}^{\phantom{\alpha}; i} \,
    \overline{\pi}_{\dot{\beta} ; j}\, ,
\end{equation}
\begin{equation}\label{betlu30b}
    y^{\alpha\dot{\beta}} =
    -
    \frac{1}{2 | f |^2} \,
    t_{i}^{\ j} \, \pi^{\alpha ;i} \,
    \overline{\pi}^{\dot{\beta}}_{\phantom{\beta} ; j} \, .
\end{equation}
\renewcommand{\theequation}{\arabic{equation}}
\setcounter{equation}{31} One gets using

\begin{eqnarray}\label{betlu31}
    \overline{\pi}^{{\dot\alpha}}_{\phantom{\alpha};i} \, \overline{\pi}_{\dot\alpha ; j}
    = - \epsilon_{ij} \, f\, ,
    \qquad
{\pi}^{{\alpha} ; i} \, {\pi}_{{\alpha}}^{\phantom{\alpha}; j}
    = - \epsilon^{ij} \, \overline{f}\, ,
    \cr\cr
      \overline{\pi}_{\dot{\alpha}; i} \, \overline{\pi}_{\dot\beta}^{\phantom{\alpha ; };i}
    = \epsilon_{\dot\alpha\dot \beta} \, f\, ,
    \qquad
{\pi}_{{\alpha} ;i} \, {\pi}_{{\beta}}^{\phantom{\beta} ; i}
    =  \epsilon_{{\alpha} {\beta}} \, \overline{f}\, ,
\end{eqnarray}
that
\begin{eqnarray}\label{betlu32}
    y^{\alpha \dot{\beta}} \,
     {\pi}_{\alpha}^{\phantom{\alpha}; k} & = &
    \frac{1}{2f} \, t_{i}^{\phantom{i} j} \,
     \overline{\pi}^{\dot{\beta}}_{\phantom{\beta}; j}
    \, \epsilon^{ik} \, ,
    \cr\cr
 y^{\alpha \dot{\beta}} \,
  \overline{\pi}_{\dot{\beta}; k} & = &
    \frac{1}{2{\overline{f}}} \, t_{i}^{\phantom{i} j} \, {\pi}^{{\alpha}; i }
    \, \epsilon_{jk} \, .
\end{eqnarray}
Substituting in (\ref{betlu29}) the formula (\ref{betlu17}) and
using
 (\ref{betlu32}) one gets $((t^{i}_{\phantom{i} j})^{*} = t^{\phantom{i} j}_{i})$

\begin{eqnarray}\label{betlu33}
\Theta & = &
 P_\mu \, dx^\mu + i
 \left[-
\frac{1}{2 \overline{f}} \, t_{i}^{\phantom{i} j} \, {\pi}^{{\alpha}
; i}
 \, \epsilon_{jk} \, d {\pi}_{\alpha}^{{\phantom{\alpha}}; k}
 + \frac{1}{2f} \, t_{i}^{\phantom{i} j} \, \overline{\pi}^{\dot\beta}_{\phantom{\alpha} ; j}
\, \epsilon^{ik} \, d \overline{\pi}_{\dot\beta ; k}
 \right]
 \cr\cr
& = & P_\mu \, dx^\mu + i
 \left[
\frac{1}{2 \overline{f}} \, t_{i}^{\phantom{i}  j} \, {\pi}^{\alpha
; i} \,  d {\pi}_{\alpha ;j}
 + \frac{1}{2{f}} \, t_{i}^{\phantom{i} j}
 \overline{\pi}^{\dot\beta}_{\phantom{\alpha} ; j} \,
 d \overline{\pi}_{\dot\beta}^{\phantom{\beta} ; i}
 \right] \, .
\end{eqnarray}

Let us observe that\footnote{We define $A_{(i } B_{ j) } =
\frac{1}{2} (A_i B_j + A_j B_i )$ and $ A_{ [i } B_{ j]} =
\frac{1}{2}( A_i B_j - A_j B_i)$. }

\begin{eqnarray}\label{betlu35new}
\overline{\pi}^{\dot{\beta}}_{\phantom{\beta} ; i} \,
 d \overline{\pi}_{\dot{\beta}; j}
& = &
 \overline{\pi}^{\dot{\beta}}_{\phantom{\beta} ;[i}\,
   d \overline{\pi}_{\dot{\beta}; j]}
 + \overline{\pi}^{\dot{\beta}}_{\phantom{\beta} ;(i}
 d \overline{\pi}_{\dot{\beta}; j )}\, ,
 \cr\cr
\pi^{{\beta}; i} \, d \pi_{{\beta}}^{\phantom{\beta} ;j} & = &
 \pi^{{\beta}; [i}\, d \pi_{{\beta}}^{\phantom{\beta} ;  j]}
 + \pi^{{\beta}; (i } d \pi_{{\beta}}^{\phantom{\beta} ; j )}\, ,
\end{eqnarray}
and observe that
\begin{eqnarray}\label{betlu36new}
\overline{\pi}^{\dot{\beta}}_{\phantom{\beta} ;[ i} \, d
\overline{\pi}_{\dot{\beta}; j]} & = & -\frac{1}{2} \,
\epsilon_{ij}\, d {f} \, , \cr\cr \pi^{{\beta}; [i} \, d
\pi_{{\beta}}^{\phantom{\beta}; j]} & = & -\frac{1}{2} \,
\epsilon^{ij}\, d \overline{f} \, .
\end{eqnarray}
We obtain
\begin{eqnarray}
\Theta & = & P_\mu dx^\mu + \frac{i}{2} \ t_{i}^{\phantom{i} j}
\left( \frac{1}{\overline{f}} {\pi}^{{\alpha} ;( i} \,
 d {\pi}_{\dot{\alpha}}^{\phantom{\alpha} ; k)}
\, \epsilon_{kj} + \frac{1}{ {f}}
 \overline{\pi}^{\dot\beta }_{\phantom{\alpha} ; (j} \,
 d \overline{\pi}_{\dot\beta; k )} \, \epsilon^{ik} \right)
 + \frac{i}{4} \, t_{k}^{\phantom{k}  k}
\, d \ln \left( \frac{\overline{f}}{f} \right)\, .\nonumber\\
\label{betlu37new}
\end{eqnarray}
In the symplectic prepotential (\ref{betlu33}) the space-time
coordinates after quantization are noncommutating. In order to
introduce symplectic form with commuting space-time coordinates
$X_\mu$ one should use the formulae (\ref{betlu27}--\ref{betlu28}).
One gets
 $(\Delta x^{\alpha \dot{\beta}}
 = \mbox{Re} \Delta \, z^{\alpha \dot{\beta}})$

\begin{eqnarray}\label{betlu34}
P_{\mu} \, d x^\mu & = & P_\mu \, dX^\mu  - P_{\alpha \dot{\beta}}
\, d (\Delta \, x^{\alpha \dot{\beta}}) \cr\cr &= & P_\mu \, dX^\mu
 - \epsilon_{3rs} \,
 P_{\alpha \dot{\beta}} \,
d\left( \frac{-1}{2 | f |^2}
  \, t_r \, \pi^{\alpha ; i}
  (\tau_s)_{i}^{\ j} \, \overline{\pi}^{\dot{\beta}}_{\ ; j}\right) \, .
\end{eqnarray}
Using the formula (\ref{betlu17}) one obtains

\begin{eqnarray}
P_\mu dx^\mu&=&P_\mu
dX^\mu+P_{\alpha\dot\beta}\frac{1}{2|f|^2}\epsilon_{3rs} t_r
(\tau_s)_i^{\ j}
d(\pi^{\alpha;i}\overline{\pi}^{\dot\beta}_{\phantom{\beta};j})\nonumber\\
&=&P_\mu dX^\mu+\epsilon_{3rs} t_r (\tau_s)_i^{\
j}\left[\frac{1}{2\overline{f}}\pi_{\alpha;j}d\pi^{\alpha;i}-
\frac{1}{2f}\overline{\pi}_{\dot\beta}^{\phantom{\beta};i}d\overline{\pi}^{\dot\beta}_{\phantom{\beta};j}\right]
\, . \label{betlu39}
\end{eqnarray}

Substituting (\ref{betlu39}) in (\ref{betlu37new}) one gets
($r=1,2,3$)

\begin{equation}\label{betlu40}
\Theta=P_\mu dX^\mu+\frac{i}{f}t_r(\tau_r)_1^{\
j}\overline{\pi}^{\dot\alpha}_{\phantom{\alpha};j}d\overline{\pi}_{\dot\alpha;2}-
\frac{i}{\overline{f}}t_r(\tau_r)_j^{\
1}{\pi}^{\alpha;j}d{\pi}_{\dot\alpha}^{\phantom{\alpha};2}+\frac{i}{2}(t_0+t_3)\left(\frac{d\overline{f}}{\overline{f}}-\frac{df}{f}\right)
\end{equation}
or
\begin{equation}\label{betlu41}
\Theta=P_\mu
dX^\mu-i\left(\overline{\sigma}^{\dot\alpha}_{\phantom{\alpha};1}d\overline{\pi}_{\dot\alpha;2}-
\sigma^{\alpha;1}d\pi_\alpha^{\phantom{\alpha};2}\right)+ed\phi\,
,
\end{equation}

where ($r=1,2,3$)

\begin{eqnarray}\label{betlu42}
\sigma^{\alpha;i}&=&-\frac{1}{\overline{f}}t_r(\tau_r)_j^{\
i}\pi^{\alpha;j}\, ,
\\
\overline{\sigma}^{\dot\alpha}_{\phantom{\alpha};i}&=&-\frac{1}{f}t_r(\tau_r)_i^{\
j}\overline{\pi}^{\dot\alpha}_{\phantom{\alpha};j}\, ,\nonumber
\\
e&=&t_0+t_3\, ,
\nonumber\\
\phi&=&\frac{i}{2}\ln\frac{\overline{f}}{f}\, ,
\nonumber\\
\end{eqnarray}

implying

\begin{equation}\label{betlu43}
d\phi=\frac{i}{2}\left(\frac{d\overline{f}}{\overline{f}}-\frac{df}{f}\right)
\,.
\end{equation}

We see therefore that the second part of the symplectic form
(\ref{betlu41}) is the second rank fixed spinor in the internal
space, and represents in the internal three-space ($SU(2)\simeq
O(3)$) the direction $a_1+ia_2$. The third part, describing the
electric charge sector, indicates the direction $a_0+a_3$ ($a_0$ is
a scalar), i.e. the $SU(2)$ symmetry is also broken. It can be
mentioned that the formula (\ref{betlu42}) for the electric charge
recalls the Gell-Mann-Nishijima formula
$Q=I_3+\textstyle{\frac{Y}{2}}$, where in $Y$ is a scalar from the
point of view of the internal $SU(2)$ isospin symmetry.

Let us point out that the 18 variables
($X_\mu,P_\mu,\pi^{\alpha;2},\overline{\pi}_{\dot\alpha;2},\sigma^{\alpha;1},\overline{\sigma}_{\dot\alpha;1},e,\phi$)
as composites of two-twistor space coordinates are not independent. On can show from
(\ref{betlu17}) and (\ref{betlu42}) (see also \cite{betl10new}\footnote{We should mention that the
constraints $R_1\cdots R_6$ are linear combinations of the constraints given in \cite{betl10new}})
that they satisfy the following three kinematical constraints:

\begin{eqnarray}\label{betlu45}
R_1&=&\sigma_\alpha^{\phantom{\alpha};1}P^{\alpha\dot\beta}\overline{\sigma}_{\dot\alpha;1}-\mathbf{t}^2=0\,,\qquad\mathbf{t}^2=(t_1)^2+(t_2)^2+(t_3)^2\\
\label{betlu49a}
R_2&=&\pi_\alpha^{\phantom{\alpha};2}P^{\alpha\dot\beta}\overline{\pi}_{\dot\beta;2}-\frac{1}{2}P^2=0
\, ,\\
\label{betlu47}
R_3&=&\pi_\alpha^{\phantom{\alpha};2}\sigma^{\alpha;1}-\overline{\pi}_{\dot\alpha;2}\overline{\sigma}^{\dot\alpha}_{\phantom{\alpha};1}=0
\, ,
\end{eqnarray}

where (we recall that $f=\sqrt{\frac{P^2}{2}}e^{i\phi}$)
\renewcommand{\theequation}{48\alph{equation}}
\setcounter{equation}{0}
\begin{eqnarray}\label{betlu46a}
t_1^{\ 2}=
t_1-it_2&=&-\frac{1}{\overline{f}}\pi_\alpha^{\phantom{\alpha};2}P^{\alpha\dot\beta}\overline{\sigma}_{\dot\beta;1}
\, ,\\
t_2^{\ 1}=t_1+it_2&=&
-\frac{1}{f}\sigma_{\alpha}^{\phantom{\alpha};1}P^{\alpha\dot\beta}\overline{\pi}_{\dot\beta;2}
\, , \label{betlu46b}\\
t_1^{\ 1}-t_2^{\
2}=2t_3&=&\pi^{\alpha;2}\sigma_\alpha^{\phantom{\alpha};1}-
\overline{\pi}_{\dot\alpha;2}\overline{\sigma}^{\dot\alpha}_{\phantom{\alpha};1}
\, . \label{betlu46c}
\end{eqnarray}
\renewcommand{\theequation}{\arabic{equation}}
\setcounter{equation}{48}

The constraints (\ref{betlu45})--(\ref{betlu47})  reduce the 18
independent variables in (\ref{betlu41}) to 16 degrees of freedom
in two-twistor space because only two of them are independent.

The symplectic form (\ref{betlu41}) implies the following canonical Poisson brackets (CPB):

\renewcommand{\theequation}{49\alph{equation}}
\setcounter{equation}{0}
\begin{eqnarray}
&& \quad \left\{ X_\mu , X_\nu \right\} = 0 \,
 ,\quad \left\{ P_\mu , P_\nu \right\} = 0 \,
 ,\quad \left\{ P_\mu , X_\nu \right\} = \eta_{\mu\nu}\,
 ,\label{betlu48a}
 \\ \cr
  && \qquad \quad \left\{ \pi_{\alpha}^{\phantom{\alpha};2} ,
 \sigma^{\beta;1} \right\} =i\delta_{\alpha}^{\ \beta}
 \, , \qquad \left\{ \overline{\pi}_{\dot{\alpha};2} ,
 \overline{\sigma}^{\dot{\beta}}_{\phantom{\alpha};1}\right\} = -i
 \delta_{\dot{\alpha}}^{\ \dot{\beta}} \, ,\\
 \cr  && \qquad \qquad \qquad
 \qquad \qquad \left\{ e, \phi\right\} = 1 \, .\label{betlu48}
\end{eqnarray}
\renewcommand{\theequation}{\arabic{equation}}
\setcounter{equation}{49}

Let us observe that the three constraints ($R_1,R_2,R_3$) have the following CPB:

\renewcommand{\theequation}{50\alph{equation}}
\setcounter{equation}{0}
\begin{eqnarray}\label{betlu48'}
\{R_1,R_2\}&=&-2i\sigma_{\alpha}^{\phantom{\alpha};1}P^{\alpha\dot\beta}\overline{\sigma}_{\dot\beta;1}
\, ,\\
\{R_2,R_3\}&=&2i\pi_\alpha^{\phantom{\alpha};2}P^{\alpha\dot\beta}\overline{\pi}_{\dot\beta;2}
\, , \\
\{R_1,R_3\}&=&-iR_2\cdot R_3\, .
\end{eqnarray}
\renewcommand{\theequation}{\arabic{equation}}
\setcounter{equation}{50}
 We suplement them with the following additional three physical
constraints:

\renewcommand{\theequation}{51\alph{equation}}
\setcounter{equation}{0}
\begin{eqnarray}
R_4&=&\mathbf{t}^2-s(s+1)=0\label{betlu49b}
\, ,\\
R_5&=&P^2-m^2=0
\, ,\\
R_6&=&e-e_0=0\, . \label{betlu49c}
\end{eqnarray}
\renewcommand{\theequation}{\arabic{equation}}
\setcounter{equation}{51}
 Using the canonical PB (see (\ref{betlu48a}-c)) one gets the following
relations: ($A,B=3,4,5,6$)

\begin{eqnarray}
\{R_A,R_B\}=0 \, .
\end{eqnarray}
We see therefore that one can consider the set ($R_3,R_4,R_5,R_6$) as four first class constraints.

The constraints $R_4=R_5=0$ can be interpreted as determining the numerical value of the mass
operator $P^2$ and the isospin square $\mathbf{t}^2$. If we observe further that
\cite{betl6,betl10new}

\begin{equation}\label{betlu52}
\mathbf{t}^2=-\frac{1}{2|f|^2}W_{\alpha\dot\beta}W^{\alpha\dot\beta}=-\frac{1}{P^2}W^2
 \, ,
\end{equation}
where $W^2\equiv W_{\alpha\dot\beta}W^{\alpha\dot\beta}$ describes the square of the composite
Pauli-Luba\'{n}ski fourvector, we see that one can identify $\mathbf{t}^2$ with the relativistic
spin square Casimir of the Poincare algebra.

It should be added that the constraints (\ref{betlu49b}-c) can be supplemented with another
relation determining the projection of isospin on the third axis.

%
%
%
%
%
%
%
%

\section{Relativistic particles with mass, spin and electric charges
  in extended space-time.}\label{relativisticparticles}

$\quad\,$Let us consider the 18 coordinates ($X_\mu,P_\mu,\eta^\alpha\equiv\pi^{\alpha;2},
\overline{\eta}_{\dot\alpha}\equiv\overline{\pi}_{\dot\alpha;2},
\sigma^\alpha\equiv\sigma^{\alpha;1},
\overline{\sigma}_{\dot\alpha}\equiv\overline{\sigma}_{\dot\alpha;1},e,\phi$) occuring in the
symplectic one-form (\ref{betlu41}) as primary ones. They shall be restricted by four constraints
$R_3=\cdots=R_6=0$. We propose the following action with four constraints introduced through
Lagrangian multipliers:

\begin{equation}\label{betlu54}
S=\int d\tau \mathcal{L}=\int d\tau \left[P_\mu \dot
X^\mu+i(\sigma^\alpha\dot\eta_\alpha-
\overline{\sigma}^{\dot\alpha}\dot{\overline{\eta}}_{\dot\alpha})
+e\dot\phi+\sum_{i=2}^{i=6}\lambda_i R_i\right]\, ,
\end{equation}
where $R_i$ ($i=3,\dots,6$) are given by the formulae (\ref{betlu47}), (\ref{betlu49b}--c). Using
(\ref{betlu48a}-c) one can show that all four constraints  are first class. One can quantize the
model by canonical quantization of the PB (\ref{betlu48a}-c). Using the standard quantization rule
$\left(i\hbar\{a,b\}\leftrightarrow [\hat a,\hat b]\right)$ one gets the following canonical
commutators:

\begin{eqnarray}\label{betlu55}
[\hat X_\mu,\hat X_\nu]=[\hat P_\mu,\hat P_\nu]=0 \, ,
\quad &
&\quad
[\hat P_\mu,\hat X_\nu]=i\hbar \eta_{\mu\nu}
\, ,\nonumber\\
\phantom{.}[\hat \sigma^\alpha,\hat
\eta_\beta]=\hbar\delta^\alpha_\beta\, ,
 \quad&&\quad
[\hat{\overline{\sigma}}^{\dot\alpha},\hat{\overline{\eta}}_{\dot\beta}]=-\hbar\delta^{\dot\alpha}_{\dot\beta}
\, ,\nonumber\\
\phantom{.}[\hat e,\hat \phi]&=&i\hbar\, .
\end{eqnarray}

We introduce the Schr\"{o}dinger representation in the extended
momentum space
$\mathcal{P}_k=(P_\mu,\eta_\alpha,\overline{\eta}_{\dot\alpha},\phi)$
(k=1,\dots,9) as follows

\begin{eqnarray}\label{betlu56}
\hat X_\mu = -i\hbar \frac{\partial}{\partial P^\mu}\, , \qquad
\hat\sigma_\alpha&=&\hbar\frac{\partial}{\partial\eta^\alpha}\, , \qquad
\hat{\overline{\sigma}}_{\dot\alpha}=-\hbar \frac{\partial}
{\partial\overline{\eta}^{\dot\alpha}}
\, ,\nonumber\\
\hat e&=&i\hbar\frac{\partial}{\partial \phi}\, .
\end{eqnarray}

In our quantized model the dynamics is characterized by the
following four wave equations describing the wave function
$\Psi(\mathcal{P}_k)\equiv\Psi(P_\mu,\eta_\alpha,\overline{\eta}_{\dot\alpha},\phi)$

\renewcommand{\theequation}{57\alph{equation}}
\setcounter{equation}{0}
\begin{eqnarray}
R_3=0:&& \left(\eta_\alpha\frac{\partial}{\partial\eta^\alpha}+
\overline{\eta}_{\dot\alpha}\frac{\partial}{\partial{\overline{\eta}^{\dot\alpha}}}\right)\Psi(\mathcal{P}_k)=0
\, ,\label{betlu57a}\\
R_4=0:&&
\left[\eta_\alpha\oeta_\dalpha\frac{\partial}{\partial\eta_\alpha}\frac{\partial}{\partial\oeta_\dalpha}
 \right.\nonumber\\
&&+ \ \frac{1}{P^2}P_{\alpha\dalpha}P_{\beta\dbeta}
\left(\eta^\alpha\epsilon^{\dbeta\dalpha}\frac{\partial}{\partial\eta_\beta}+\oeta^\dbeta\epsilon^{\alpha\beta}\frac{\partial}{\partial\oeta_\dalpha}+
2\eta^\alpha\oeta^\dbeta
\frac{\partial}{\partial\eta_\beta}\frac{\partial}{\partial\oeta_\dalpha}\right)
 \nonumber\\
&&+\ \left.\frac{s(s+1)}{\hbar^2}\right]\Psi(\mathcal{P}_k)=0
\, ,\label{betlu57c}\\
R_5=0:&& \left(P^2-m^2\right)\Psi(\mathcal{P}_k)=0
\, ,\label{betlu57b}\\
R_6=0:&& \left(\frac{\partial}{\partial
\phi}+\frac{i}{\hbar}e_0\right)\Psi(\mathcal{P}_k)=0
\, .\label{betlu57d}
\end{eqnarray}
\renewcommand{\theequation}{\arabic{equation}}
\setcounter{equation}{57}

The set (\ref{betlu57a}-d) describes the quantized first class constraints. If we add fifth
constraint $R'_4=t_3-m_3=0$, where $-s\le m_3\le s$, we obtain the description of a massive
relativistic particle with spin $s$, isospin projection $m_3$ on the third axis and the electric
charge $e_0$. One can observe that the differential form of the Lorentz-invariant third component
of isospin $t_3$
\begin{equation}
t_3=\frac{1}{2}\left(\eta_\alpha\frac{\partial}{\partial\eta_\alpha}-\oeta_\dalpha\frac{\partial}{\partial\oeta_\dalpha}\right)
\end{equation}
corresponds to the twistorial helicity formula for massless particles \cite{betl1}-\cite{betl4}
employed recently in Witten's tensorial string theory \cite{betl21new}

It should be stressed that in distinction with \cite{betl10new} in our model we do not have the
constraints $R_1=R_2=0$, i.e. our phase space can not be identified with the two-twistorial phase
space. Let us observe, however, that the conditions $R_1=0$ or $R_2=0$ can be obtained as
gauge-fixing conditions for the gauge freedom generated by the constraint $R_3$, and viceversa, the
transition from second to first class constraints is obtained by so-called gauge unfixing procedure
(see e.g. \cite{betlm17new,betlm18new}).

More detailed discussion of the quantization of the model (\ref{betlu54}) will be given in a
subsequent publication \cite{betl14new}.

\section{Final Remarks}

$\quad\,$ In the present lecture it is described the ``physical basis'' for two-twistor phase space
and considered the particle models based on the symplectic form (\ref{betlu41}).

We  stress that for the description of massive relativistic particles with spin we consider both
relativistic phase space coordinates ($X_\mu,P_\mu$) as composite (compare e.g. with
\cite{betlm20new} where the coordinates $X_\mu$ are elementary) as well as we do not introduce any
additional degrees of freedom besides two-twistor space (compare with \cite{betlm21new,betlm22new}
where additional so-called index spinor was introduced).

In comparison with the results given in \cite{betl10new} we presented here the following two new
aspects:

i) In the process of introduction of a ``physical basis'' in two-twistor space, defining the
enlarged coomposite relativistic phase space $\mathcal{M}^{18}$, we exhibited explicitely the
covariance and the breaking of the internal symmetry $SU(2)$.

ii) We introduced the particle model only inspired by the two-twistor space geometry with entirely
first class constraints. If we wish to link the phase space of our model with two-twistor manifold
and composite twistor formulae for
$P_\mu,X_\mu,\eta_\alpha,\overline{\eta}_{\dot\alpha},\sigma_\alpha,\overline{\sigma}_{\dot\alpha},
e$ and $\phi$ one should consider the gauge-fixed version of the model. In such a way will appear
the second class constraints, considered in \cite{betl10new}.

In this lecture there is presented only the model in four dimensions ($D=4$). It appears that one
can extend our considerations to the pair of super-twistors (see e.g. \cite{betl15new,betl16new})
and consider the corresponding superparticle models with mass and superspin. Other generalization
consists in the extension of our discussion to other dimensions $D$ (e.g. $D=10$ or $D=11$; see
e.g. \cite{betl16new}-\cite{betl20new}).

\medskip
\subsection*{Acknowledgments}
We would like to thank J.A. de Azc\'{a}rraga, S. Fedoruk and A. Frydryszak for valuable remarks.
One of the authors (J.L.) would like to acknowledge the support of the KBN grant 1 P03B 01828 and
one of us (C.M-E.) wishes to thank the Spanish M.E.C. for his research grant.

\end{document}